\documentclass{mem}
\usepackage{natbib}\usepackage{txfonts}\usepackage{balance}
\usepackage{graphicx}
\usepackage[a4paper]{hyperref}
\idline{75}{282}
\begin{document}
\def\teff{$T\rm_{eff }$}
\def\kms{$\mathrm {km s}^{-1}$}

\title{
Using the Tremaine-Weinberg method to measure pattern speeds from H$\alpha$ velocity maps.
}

   \subtitle{}

\author{
J. E. Beckman\inst{1,2} 
\and K. Fathi\inst{1,3}
\and N. Pi\~nol\inst{1,3,4}
\and O. Hernandez\inst{5}
\and C. Carignan\inst{5}
\and I. P\'erez\inst{6}
          }

  \offprints{J.E. Beckman}

\institute{
IAC, C/ V\'\i a L\'actea s/n, Tenerife, Spain\email{jeb@iac.es}
\and
Consejo Superior de Investigaciones Cient\'\i ficas, Spain
\and
Department of Astronomy, AlbaNova University Centre, Stockholm University, Sweden
\and
Departmento de F\'\i sica, Universidad de Murcia, Spain
\and
LAE, Universit\'e de Montr\'eal, Montr\'eal, QC, Canada H3C 3J7
\and
Departamento de F\'\i sica Te\'orica y del Cosmos, Universidad de Granada, Spain
}

\authorrunning{Beckman et al.}

\titlerunning{Pattern speeds from 2D velocity maps in H$\alpha$}

\abstract{
The Tremaine-Weinberg method is a well-known model independent technique for measuring density wave pattern speeds in spiral galaxies. Here we show how it can be applied to the data cubes (maps of surface brightness and velocity) obtained in H$\alpha$ emission using a Fabry-Perot spectrometer. One of the main difficulties, the discontinuity of the H$\alpha$ emission, is resolved using the neighbouring stellar continuum delivered by the data cube. We argue from symmetry that the motions not associated with the density wave should cancel. We show that our pattern speeds are reasonable by computing corotation radii, and comparing them to measured bar lengths. Simulations including star forming gas also add credibility to our results. Nevertheless it will be necessary to compare them with results using the spectra of the stellar components to quantify any systematic deviations from valid pattern speed values.

}
\maketitle{}

\section{The ``traditional'' Tremaine-Weinberg method}
It is now well-known that, using a combination of radial velocity and surface brightness measurements along axes parallel to the major axis of a galaxy disc, it is possible to derive the pattern speed of the density wave. The method, proposed by Tremaine \& Weinberg (1984) is predicated on three conditions: (a) the disc is flat, (b) the disc has a well defined pattern speed, and (c) the surface brightness of the tracer obeys the continuity equation. The method was formulated in a straightforward way by Merrifield \& Kuijken (1995), who expressed the pattern speed $\Omega_p$ as a ratio of the integrated line of sight velocity and the integrated value of the distance along a the measurement axis, both weighted by the luminosity of the emitter at each point, finally deprojected using the measured inclination angle of the disc. To improve the significance of the measurement it can be repeated along more than one parallel axis, and the results averaged, see e.g. Corsini et al. (2003, 2007). The advantage of this method is that it is largely model independent. Its disadvantage is that it requires continuity for the component whose brightness distribution weights the velocity and distance parameters. This was the reason given by Tremaine \& Weinberg (TW) for being unable to obtain a useful pattern speed for NGC\,5383 using HI 21cm data. So the method's utility appeared confined to early type objects, where S:N considerations mean that it has been applied to relatively few objects so far. However, Zimmer et al. (2004) were able to apply it usefully to CO emission line maps, which gives a lead towards a possible application in H$\alpha$, and the extension of the method to late type galaxies.

\begin{figure*}[t!]
\center
\resizebox{\hsize}{!}{\includegraphics[clip=true]{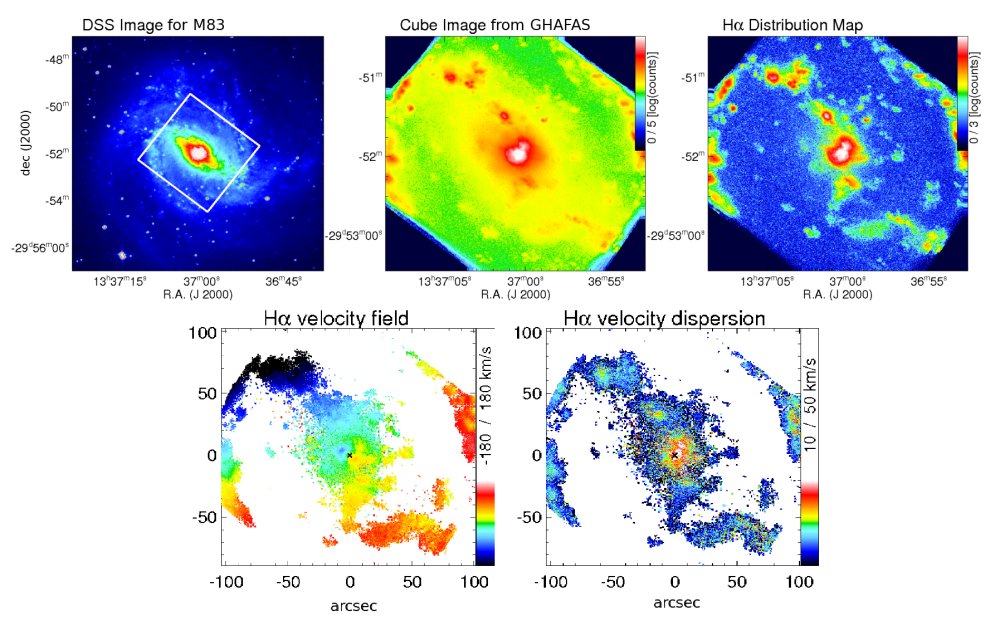}}
\caption{\footnotesize An example of the output of GHaFaS for the galaxy M\,83. All panels are labled accordingly, and at the bottom right, the inflowing gas feeding the circumnuclear starburst is outlined (Fathi et al. 2008).}
\label{m83}
\end{figure*}

\section{GHaFaS and FaNTOmM}
We developed the scanning Fabry-Perot (FP) spectrometer GHaFaS to make two dimensional velocity maps of galaxies, and one of the points we wished to test was whether we could make use of these to carry out Tremaine-Weinberg analysis and derive pattern speeds. The obvious advantage here is the full coverage of the galaxy in radial velocity (given a sufficient surface density of massive star formation) in short observing times. The disadvantage is the difficulty of measuring a surface brightness parameter whose spatial continuity is assured. First we offer a brief description of the instrument, and then explain how we went about measuring pattern speeds.

GHaFaS is installed at the Nasmyth focus of the 4.2m William Herschel Telescope (WHT) on La Palma. It has been described in detail in Hernandez et al. (2008), and some of its first results, revealing clearly the spiral inflow of gas to the nucleus of M83 were presented in Fathi et al. (2008). This instrument marks a technological advance on its predecessor FaNTOmM (Hernandez et al. 2003) with which some of the data used for the present article were obtained. Optically they are in almost all ways equivalent, and the reader is referred to the two articles just cited for all details. These spectrometers can map an extended source in a single emission line over a field of several minutes of arc, with seeing limited angular resolution of order 1 arcsec and velocity resolution a little better than 10 \kms, yielding a full velocity map for a galaxy which fits the field in a time of some two to three hours. An example of such a mapping process is shown in Fig. 1.

We can see that the velocity map is very well suited to the TW method, because  it has a large sample of points with well defined velocities. However there are two problems to be faced. Firstly we need a surface brightness map from a component which obeys the continuity equation. We can see that the H$\alpha$ map is obviously patchy, and physically we do not expect continuity to be obeyed by the ionized component, since the geometry of the ionized zones is determined partly by the gas distribution (which in any case does not obey continuity) and partly by the distribution of ionizing stars within the gas. Secondly the velocity field, although globally dominated by the large scale gravitational potential, can be dominated locally by specific effects such as the flows around massive star forming regions, and large scale shocks. We will see how these problems are handled in the next section.

\section{Why can TW method work with FP H$\alpha$ data}
The way we tackle the continuity problem is to make use of the fact that in a data cube from the FP for each pixel of the image we have a spectrum which contains the full profile of the H$\alpha$ emission line and a sample of the neighbouring stellar continuum. The latter is a source which samples the underlying stellar population and can therefore be expected to obey continuity (at least to the same degree as the light available in the traditional mode using stellar absorption spectroscopy and a separate continuum image). Furthermore, the spatial sampling of this continuum image is identical with that of the emission line image, which is a methodological advantage. Although the signal per velocity channel in the continuum is not high, we can in general integrate over a few tens of channels to obtain a reasonable S:N ratio.

There is no ready way to avoid gas motions which do not correspond to the density wave pattern. These can, however, be considered in two main categories, global: strong shock motions associated with the bar, plus streaming motions associated with the arms, on the one hand, and local: the expanding features due to winds and supernovae around OB associations on the other. If the galaxy is a bisymmetric barred spiral the symmetry in the global motions should essentially cause their effects to cancel in the TW integrations. Each local motion is close to isotropic in the plane of the galaxy, and the distribution of the HII regions associated with the OB associations shares the global symmetry of the galaxy. Under these conditions we would not expect their integrated effects to cause major perturbations to the TW integrals, given that the axes subtend equal lengths on each side of the minor axis. To summarize, we have reasonable grounds to believe that the TW method used on this kind of data can give valid results. We will in the end be able to test these using the results of simulations, and comparing these results for a given galaxy with those from conventional stellar slit spectroscopy.

\begin{figure*}[t!]
\center
\resizebox{\hsize}{!}{\includegraphics[clip=true]{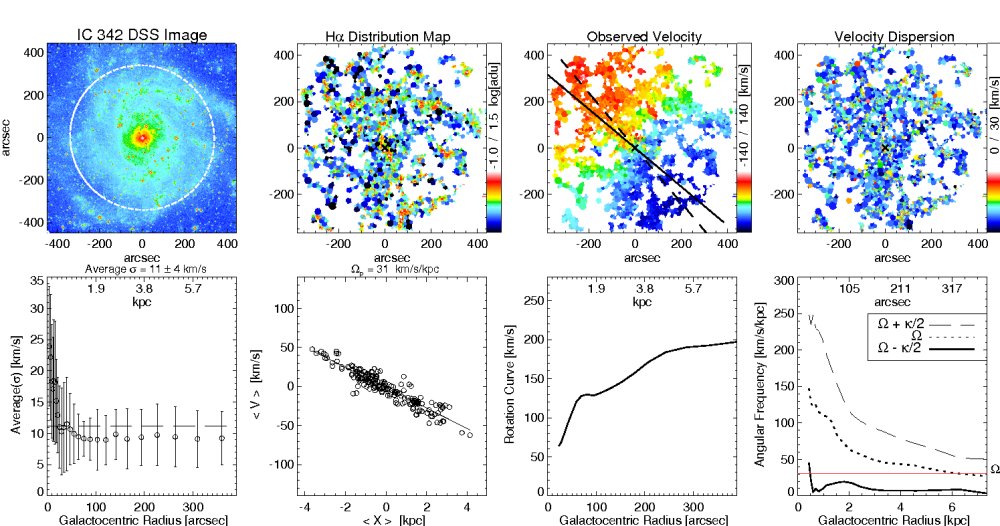}}
\resizebox{\hsize}{!}{\includegraphics[clip=true]{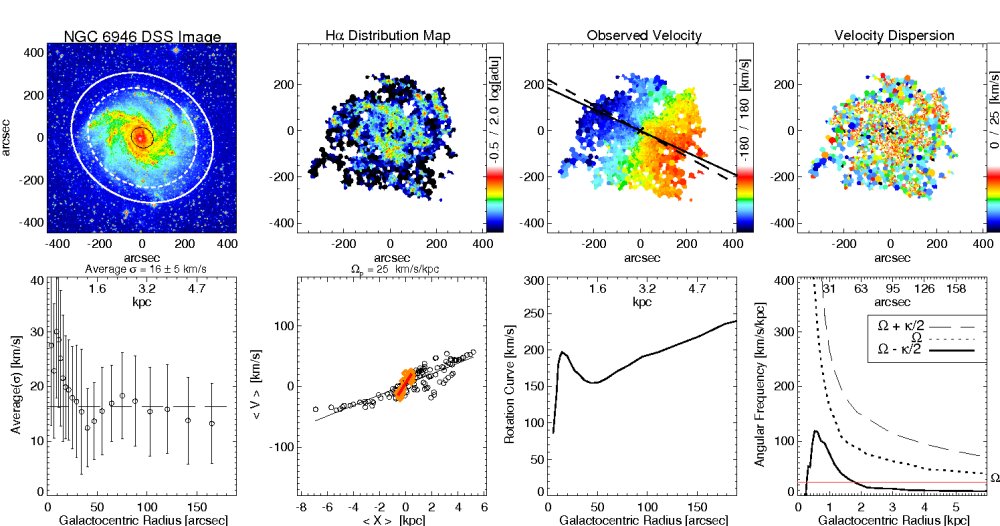}}
\caption{\footnotesize Output products of the FP mapping process which can be used for kinematic analysis of galaxies, shown for IC\,342 and for NGC\,6946.For each object we present (left to right, first top row and then bottom row: DSS red image; H$\alpha$ surface brightness map (0th moment), velocity map (1st moment), velocity dispersion map (2nd moment), radial variation of azimuthally averaged dispersion, the TW plot (see text), the rotation curve, and resonance analysis curves derived from the latter.}
\label{examples}
\end{figure*}

\begin{figure*}[t!]
\resizebox{\hsize}{!}{\includegraphics[clip=true]{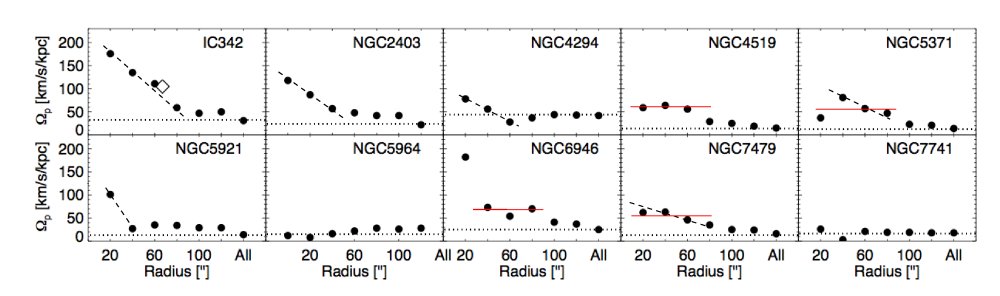}}
\caption{\footnotesize Plots of derived pattern speed against maximum radius of disc used, for all the galaxies of our sample. Dotted line: Asymptotic value at large radius, giving our derived value of $\Omega_p$. Dashed line: Declining, spurious values of $\Omega_p$ from zone of smaller radius than the bar length. Solid ``red'' line. Where the presence of a secondary bar implies a possible secondary pattern speed this is indicated (where short and long dashed lines are superposed, we assume that there is no genuine second pattern speed,( i.e. in NGC\,5371 \& NGC\,7479).}
\label{examples}
\end{figure*}

\section{TW analysis in H$\alpha$}
In Fig. 2, we show two characteristic data sets and the elements of their analysis. As well as the data cube products we plot four derived quantities: the radial variation of the mean velocity dispersion, the TW plot (described in more detail shortly), the rotation curve, and the radial angular frequency plots. Each point on the TW plot is the result of plotting the integral of the luminosity weighted radial velocity against the integral of the luminosity weighted distance along an axis parallel to the major axis, i.e. one point per axis. The FP velocity map allows us to choose many tens of axes, and the linear fit to these plotted points gives us an accurate measure of the pattern speed, always assuming that we are applying the method correctly.

We want only to bring out two points here. The number of data points is between one and a half and two orders of magnitude greater than that available using stellar spectra and broad band imaging to derive the pattern speed, and in the case of NGC\,6946 we find two slopes on the TW plot which we can tentatively identify with the pattern speeds of the outer and inner bar respectively (although in the latter case there are good theoretical reasons to be cautious about whether the value derived corresponds to a true pattern speed). The potential power of the method is nicely illustrated in both of these plots. It is worth noting that the data set allows us to make a series of valuable tests. One is to vary the assumed inclination angle, using a value obtained from the geometry of the outer disc ellipses in broad band images, or a value obtained from the kinematics measured here. We found that using the morphological position angle led to TW plots with lower scatter, which is not surprising since the kinematic plots include the kinds of variant velocities discussed above. We could optimize the inclination angle by minimizing the scatter, with a characteristic angular uncertainty of some 2 degrees. The second is to vary the fraction of the disc over which the TW axial lengths were estimated. We found that the scatter in the plots was minimized using the maximum disc cover, out to the largest radius with significant signal, which is as predicted in Tremaine \& Weinberg (1984). A fuller description of these pieces of work can be found in Fathi et al. (2009), and in Table 1 we will just summarize the final results of applying the method to all the galaxies analyzed. It is clear that the values of the pattern speed obtained using H$\alpha$ and the continuum for the luminosity weighting are in good agreement, so that the continuity problem for H$\alpha$ does not, in the end, appear to be as serious as expected.

\begin{table}
\caption{Comparison of the pattern speed values obtained using the TW method, weighting the velocities and positions with the Ha surface brightness (middle) and with the neighbouring continuum stellar brightness (right). The agreement is quite good, within the errors, but we consider the continuum weighted values as more reliable, since continuity holds more closely. The continuum comes from the same FP cube as the H$\alpha$.}
\label{abun}
\begin{center}
\begin{tabular}{lcc}
\hline
Galaxy & $\Omega_p$ (H$\alpha$) & $\Omega_p$ (Continuum)\\[1mm]
\hline\hline
IC\,342   &  31$_{-1}^{+5} $& 32    \\[1mm]
NGC\,2403 &  22$_{-1}^{+6} $& 25    \\[1mm]
NGC\,4294 &  44$_{-10}^{+3}$& 38    \\[1mm]
NGC\,4519 &  15$_{-2}^{+3} $& 22    \\[1mm] 
NGC\,5371 &  14$_{-1}^{+5} $& 15    \\[1mm]
NGC\,5921 &  14$_{-2}^{+2} $& 11    \\[1mm]
NGC\,5964 &  25$_{-5}^{+1} $& 20    \\[1mm]
NGC\,6946 &  25$_{-6}^{+6} $& 27    \\[1mm]
NGC\,7479 &  16$_{-2}^{+3} $& 12    \\[1mm]
NGC\,7741 &  18$_{-2}^{+13}$& 18    \\[1mm] \hline 
\end{tabular}
\end{center}
\end{table}

\begin{figure*}[t!]
\begin{center}
\includegraphics[width=14cm]{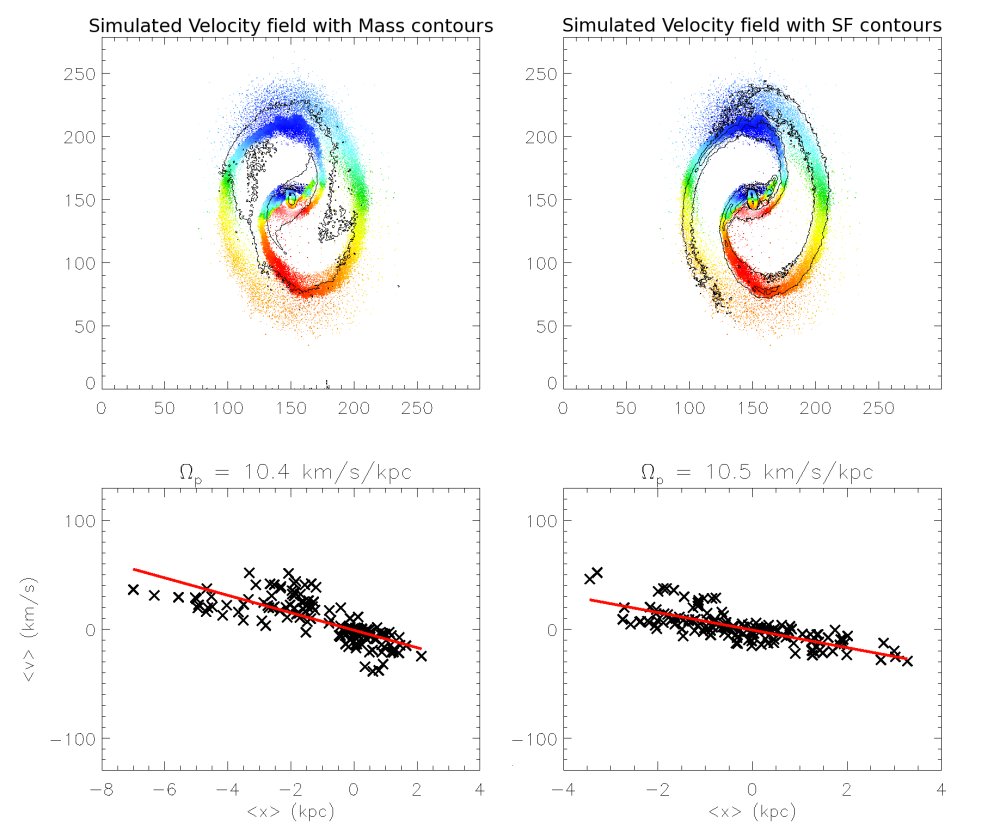}
\end{center}
\caption{\footnotesize Results of a simulation, using the stellar component to give the weighting in the TW calculation, and the stellar velocity field. The derived pattern speed converges as an increased fraction of the disc is used. The resulting pattern speed is very similar to that computed using the star formation surrogate brightness and velocity distribution. The principal component of the difference is due to the lack of continuity in the star formation. }
\label{examples}
\end{figure*}

\section{Simulations, theoretical considerations}
Our first check that the pattern speeds obtained have plausible values mas made by using them to derive the co-rotation radii for the density wave pattern in the standard way, and carefully deriving the bar lengths for comparison, using three independent techniques, of which the most consistent was to find the radius at which the ellipticity shows a clear drop. Our results gave co-rotation radii between 1 and 1.3 times the bar length, with errors of order 0.2 bar lengths. This test was designed only to show that the range of values is plausible, (and not, at this stage, to distinguish slow from fast bars) which is a result we can thus confirm. The second test was to simulate a barred galaxy, with dark matter, stars, and star forming gas, and to ``measure'' its pattern speed using the velocities of the star forming gas. The main point here was to see whether the observational result that, using only the central part of a galaxy disc well within the bar radius, we obtain artificially high values for $\Omega_p$, is reproduced in the simulations. We found this to be the case (as would be expected on rather general theoretical grounds). In Fig. 3 we show the results of this procedure, i.e. using an increasing fraction of the disc to obtain a convergent value for the pattern speed, as used on our observational data. The figure caption explains how the data are used to find the resulting pattern speed (or speeds).

The simulated galaxy shown in Fig. 4 gives a qualitatively similar result, which strengthens our view that we are obtaining plausibly valid pattern speeds using the FP data cubes. However a final observational test will be to select galaxies from the list in Table 1, observe them using the conventional long slit method with absorption spectra from the stellar population, and compare the pattern speeds so derived. If we find good agreement this will establish the FP method as a practical tool, capable of giving precise results thanks to the large number of  good quality data points made available across the galaxy disc. 

The use of a scanning FP spectrometer to measure the two-dimensional velocity field of the ionized gas appears to be a promising way to derive pattern speeds using the Tremaine Weinberg method. The observational advantages are demonstrated here, and we await corroborative tests with slit spectroscopy of the stellar component before launching a major programme of H$\alpha$ observations.

\begin{acknowledgements}
This research was supported by grant AYA2007-67626-CO2-01 of the Spanish Ministry of Education and Science.
\end{acknowledgements}

\bibliographystyle{aa}

\end{document}